\documentclass[twocolumn,showpacs,preprintnumbers,amsmath,amssymb]{revtex4}
\usepackage{graphicx}
\usepackage{dcolumn}
\usepackage{bm}
\usepackage{subfigure}

\begin{document}
\date{\today}
\title{The fermi arc and fermi pocket in cuprates  in a short-range diagonal stripe phase}
\author{W. LiMing}\email{wliming@scnu.edu.cn}
\author{Sha Ke, Jiayun Luo, Chengping Yin, Liangbin Hu}
\affiliation{Dept. of Physics, and Laboratory of Quantum Information Technology, School of
Physics and Telecommunication Engineering,  South China Normal University, Guangzhou
510006, China}
\begin{abstract}
In this paper we studied the fermi arc and the fermi pocket in cuprates in a short-range diagonal stripe phase with wave vectors $(7\pi/8, 7\pi/8)$, which reproduce with a high accuracy the positions and sizes of the fermi arc and fermi pocket and the superstructure in cuprates observed by Meng et al\cite{Meng}.  The low-energy spectral function indicates that the fermi pocket results from the main band and the shadow band at the fermi energy. Above the fermi energy the shadow band gradually departs away from the main band, leaving a fermi arc. Thus we conclude that the fermi arc and fermi pocket can be fully attributed to the stripe phase but has nothing to do with pairing.  Incorporating a d-wave pairing potential in the stripe phase the spectral weight in the antinodal region is removed, leaving a clean fermi pocket in the nodal region.
\end{abstract}
\maketitle

\section{Introduction}

The pseudo-gap (PG) state of cuprate superconductors in the normal state is still a mystery to the mechanism of  high Tc superconductivity. It opens at temperature $T^*$ , which is much higher than the transition temperature Tc of superconductivity. The fermi surface of an underdoped cuprate in the PG phase is suppressed in the antinodal region (around ($\pm \pi, 0$) and ($0, \pm\pi)$) and shows a residue part (fermi arc) around the nodal point$(\pm\pi/2, \pm \pi/2)$ \cite{Doiron,Banura}. In the overdoping region, however, this phenomenon disappears and a full fermi surface recovers. The fermi arc is usually enclosed by a weak pocket (fermi pocket). This phenomenon has been strongly confirmed by  a series of measurements of the angle-resolved photo-emission spectroscopy (ARPES)\cite{Meng,Tanaka,Lee} and some other techniques.
Since the PG is similar to a d-wave gap which opens at the antinodal region and shrinks to zero at the nodal point, it was thought ever as the same gap as  the d-wave superconducting gap. This is the one-gap scenario of the PG state, where electrons (or holes) are pre-paired incoherently above Tc. More and more evidences, however, showed that the PG is different from the superconducting gap, but a new gap resulting from such as spin modulation, charge modulation, or spin fluctuation, etc\cite{Wen,Zheng}. For instance, a few years ago Wen et al found two different energy scales in the PG state\cite{Wen}, and recently Takeshi et al found electrons and holes cannot be pre-paired at such temperatures much higher than Tc\cite{Takeshi}.

Meng et al observed rich structures of the PG state in superconductor $Bi_2Sr_{2-x}La_xCuO_{6+\delta}$ by ARPES, e.g., the Umklapp processes, the shadow bands, and the intrinsic excitations\cite{Meng}. They found  a unique superstructure wave vector $(0.24\pi, 0.24\pi)$ between the fermi pockets and the positions and sizes of the fermi pockets are almost the same at different doping level in the underdoping region. In addition, it was observed that fermi arcs and fermi pockets coexist. The understanding to these phenomena is still very controversial. Recently King et al claimed that the superstructure, the fermi pocket and the fermi arc found by Meng et al are not intrinsic but result from a surface reconstruction\cite{King}. Meng et al, however, disagreed and refuted them\cite{Zhou}.

To explain the fermi arcs and fermi pockets many models have been proposed, such as the d-density wave\cite{Oganesyan}, the d-wave check-board order\cite{Agterberg,Hoffman} , the bosonic fluctuation\cite{Han},  the phase fluctuations\cite{Qiang}, and the spin density wave (SDW) and charge density wave(CDW) stripe phase\cite{Granath}, etc.  These models reproduced the fermi arcs or fermi pockets but are  controversial to each other and contain some inner problems. For example, the d-density wave model violates the translational and rotational symmetries; the bosonic fluctuation model assumes a truncated hole band and lacks a clear physical picture; and the phase fluctuation model produces a fermi arc without a clear cutoff from the antinodal region and creates no fermi pocket.

In the last two decades one-dimensional transverse charge and spin stripes were observed by neutron scattering experiments and the measurements on the Nernst and Seebeck Coefficients\cite{Abbamonte,Mook,Tranquada,Hinkov,Haug,Chang}. Various stripe phase models were considered to explain the magnetic susceptibilities and other properties of cuprate superconductors\cite{Seibold, Matthias}.  A recent work done by Granath and Andersen employed a transverse stripe phase to model the fermi arcs and fermi pockets in cuprates\cite{Granath}. Their fermi pocket, however, is too small and asymmetric about the nodal point, disagreeing with observations. This is the main motivation of the present paper. In this paper we try to explain the superstructure and fermi arc observed by Meng et al  as a short-range diagonal stripe phase with wave vectors $(7\pi/8, 7\pi/8)$.  Although a diagonal stripe phase was only observed under doping level 5.5\% a short-range diagonal stripe phase may still exist in cuprates. The positions and sizes of the fermi pockets  can be reproduced with a high accuracy by means of this diagonal stripe phase.  The superstructure found by Meng et al can also be explained in this model.

\section{A half-filled square lattice}
We first consider a two dimensional square lattice in the half-filled state (one electron per site) with an anti-ferromagnetic (AFM) magnetization.
The Hamiltonian is given as follows
\begin{align}H &= H_0 + H_{SDW}\label{hamil}
\end{align}
where $H_0$ is the free electron Hamiltonian given by $H_0= \sum_{k\sigma}
(\epsilon_{k}-\mu) C^\dagger_{k\sigma} C_{k\sigma}$ with free electron energy
$\epsilon_k= -2t(\cos k_x + \cos k_y) - 4t'\cos k_x \cos k_y$. $H_{SDW}$ is the spin-density-wave (SDW)
coupling between electron spins and the AFM magnetization written as
\begin{align}H_{SDW}=\sum_{i} {\bf M}_i \cdot {\bf S}_i
\end{align}
where ${\bf M}_i  = {\bf M}_0 \cos({\bf Q\cdot r}_i)$ is the magnetization strenghth with
a diagonal wave vector ${\bf Q}=(\pi, \pi)$  at lattice site ${\bf r}_i$, and ${\bf S}_i = C^\dagger_{i\sigma} \tau_{\sigma\rho}C_{i\rho}$  the electron
spin operator.

The Hamiltonian (1) can be exactly diagonalized into two bands in the folded Brillouin zone (fBZ) of the lattice. The two band energies are given by
\begin{align} E_\pm = {\epsilon_k+\epsilon_{k+Q}\over 2}  \pm \sqrt{({\epsilon_k-\epsilon_{k+Q}\over 2})^2+({M_0\over 2})^2}
\end{align}

To view the fermi surface of these two bands is intriguing. When $t'=0$ we obtain a clean gapped AFM insulator with a gap $M_0/2$  locating at the boundary of the fBZ.
In the case of $t' \ne 0$, however,  the gap appears at different energies  in the fBZ, so that the two bands mix each other but never touch at any position in the fBZ as seen in the inset in Fig.1(a).
The fermi surfaces for the free electron case and the SDW case with $M_0 = 0.5$ are shown in Fig.1(a). In the free electron case $M_0 = 0.0$ the red and blue curves connect each other to become a complete fermi surface.  When an SDW coupling  is built up a few parts of the fermi surface are missing, as seen as the black and cyan curves. Comparing with the two energy bands it is seen that these missing parts take place just inside the energy gap, i.e,  the fermi surface disappears when it passes  the energy gap, see the inset in the figure. One finds eight pieces of a fermi surface, which are just the so-called {\it fermi arcs}.  When the magnetization becomes further stronger (the gap increases) the fermi arcs around the four corners of the fBZ retreat gradually away, keeping the fermi arcs around the nodal point  but being shortened. This band structure character is similar to that found  by Yang {et al.}\cite{Yang} in ${\rm Bi_2Sr_2CaCu_2O_{8+\delta}}$, where a gap appears above the fermi surface in the nodal region and drops under the fermi surface in the antinodal region.

The spectral function $A(k,\omega) = - {1\over \pi} Im G(k,\omega+i\eta)$ of this band structure at zero energy $\omega=0.0$ is shown in Fig.1(b), which displays the fermi surface in the nodal and antinodal regions. It is seen that the spectral function has a cutoff between the nodal and the antinodal regions to form a fermi arc with a clean edge. In addition, a weak fermi pocket appears connecting the fermi arc in the nodal region. The two back sides of the fermi pocket are clearly symmetric about the fBZ boundary. It is quite similar to that observed in cuprates in 90's by Marshall et al\cite{Marshall}. In fact, this pocket spectrum can be easily understood from the fermi surface shown in Fig.1(a). Mapping the energy bands in the fBZ  into the extended Brillouin zone(BZ) (unfolded), one finds a main band (MB) and a shadow band (SB) (see Fig.2a).  Including the occupation probability of the Fermi-Dirac function one obtains the weak side and the strong side of the fermi pocket.
Hence, there is  a close relation between the fermi arcs and  the band structure of a SDW state. The cutoff of the fermi surface is in fact a nesting effect with a wavevector  $(\pi, \pi)$ on the fermi surface. Further computations show that a SDW state of underdoped cuprates creates similar fermi arcs and fermi pockets.

\begin{figure}
\subfigure{\includegraphics[width=4.4cm,height=3.8cm]{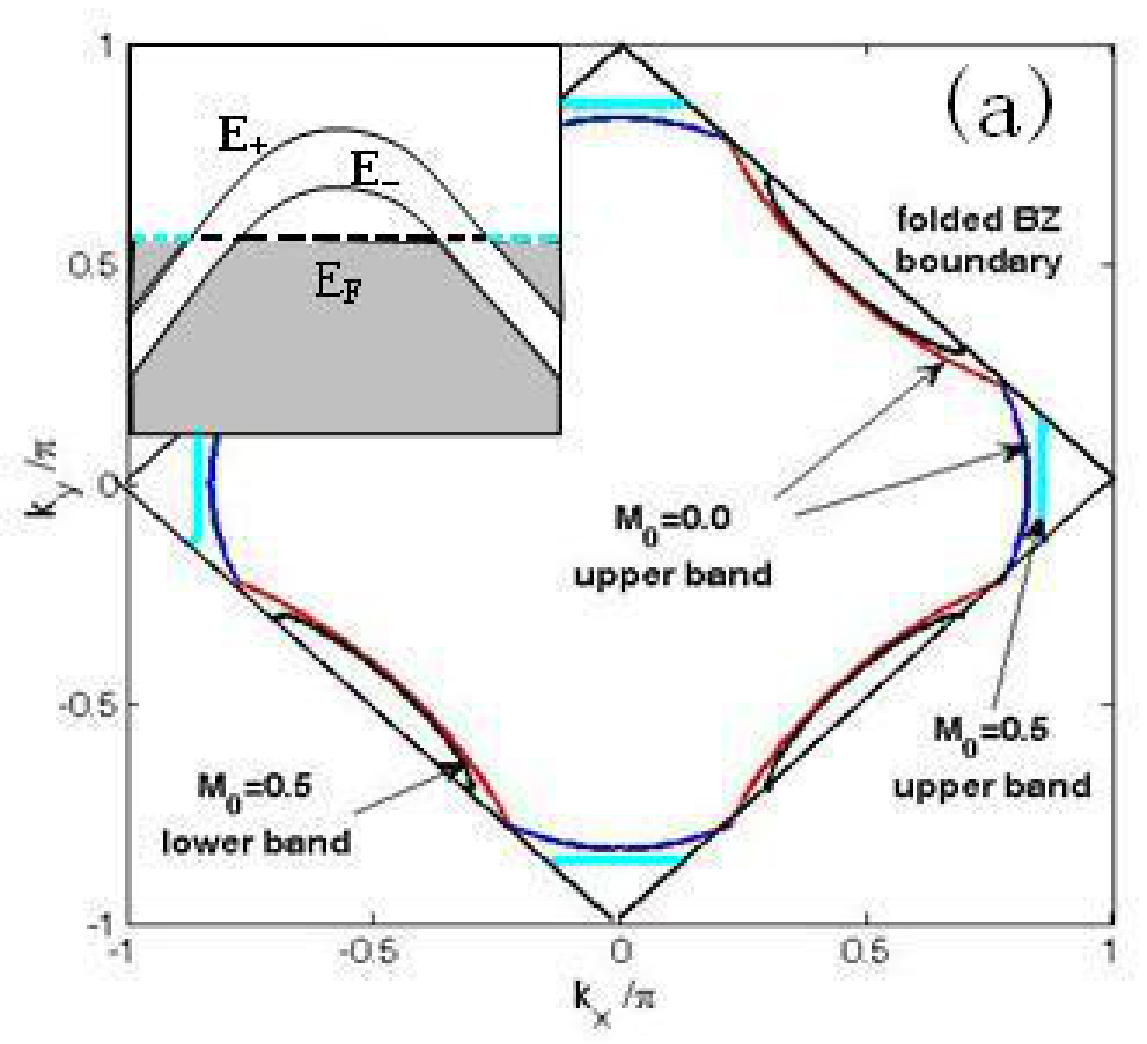}}
\subfigure{\includegraphics[width=4.4cm,height=3.8cm]{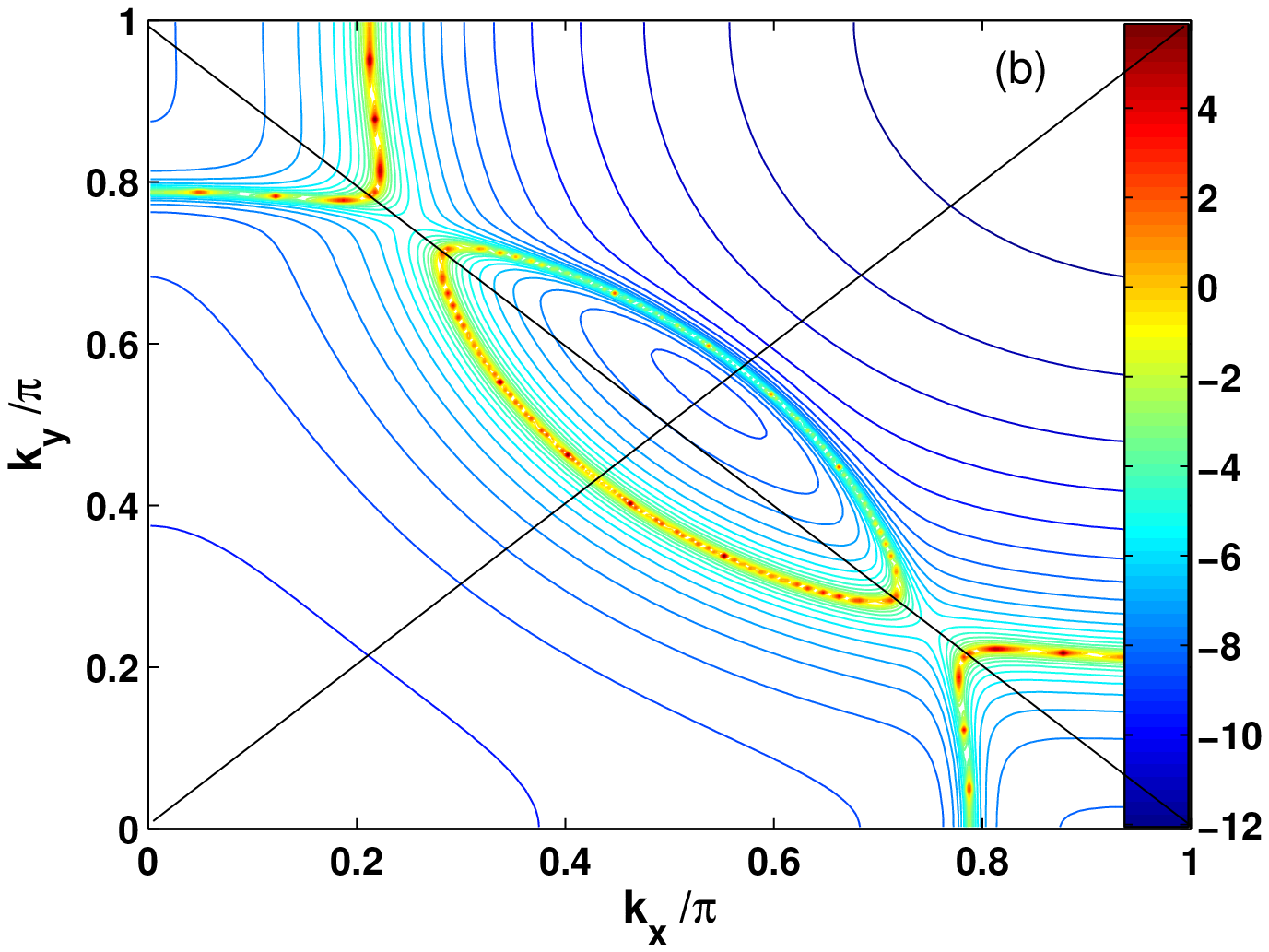}}
\caption{(a)The fermi surfaces of a square lattice without doping  in the free electron state $M_0 = 0.0$ (red and blue curves) and the SDW state $M_0 = 0.5$ (black and cyan cures). The inset shows schematically the two band energies, $E_+$ and $E_-$, on the folded BZ boundary and the fermi energy $E_F$ in the SDW state $M_0 = 0.5$.
(b)The logarithmic spectral function  $A(k,\omega=0.0)$ in the SDW state in the first quadrant. The band parameters are set to be $t =1, t' = -0.4, \mu = -0.8, kT= 0.002$ .}
\end{figure}
\section{The pseudogap state of an underdoped cuprate}

Granath and Andersen proposed a stripe model for the pseudogap state in cuprates, where one-dimensional transverse stripes with an 8-site periodicity  produce fermi pockets around the nodal region\cite{Granath}. Their fermi pocket, however, is not symmetric about the nodal point, and has a too small size and wrong position in the BZ, not agreeing with observations.  They cannot explain the diagonal superstructure with wave vector $(0.24\pi, 0.24\pi)$ observed by Meng et al.  To model this fermi pocket we consider a diagonal stripe phase with wave vector $Q = (7\pi/8, 7\pi/8)$. Inelastic neutron scattering has revealed that dynamical magnetic correlations change from a diagonal incommensurate phase to  a commensurate phase in ${La_{2-x}Sr_xCuO_4}$ in a low doping level ($0.01< x < 0.06$). A short-range diagonal stripe phase may still exist above this doping level.

\begin{figure}
\subfigure{ \includegraphics[width=4.2cm,height=3.8cm]{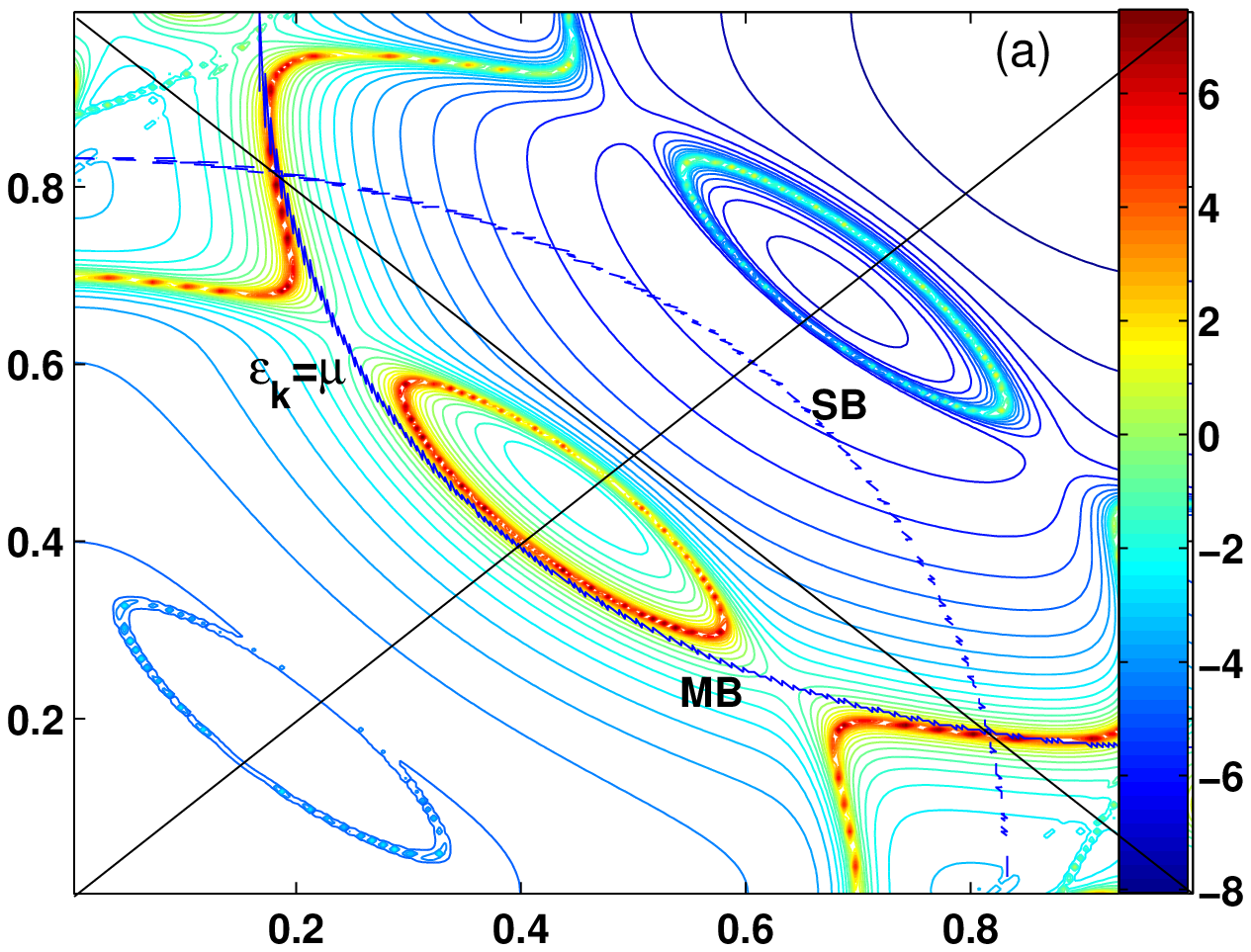}}
\subfigure{ \includegraphics[width=3.7cm,height=3.7cm]{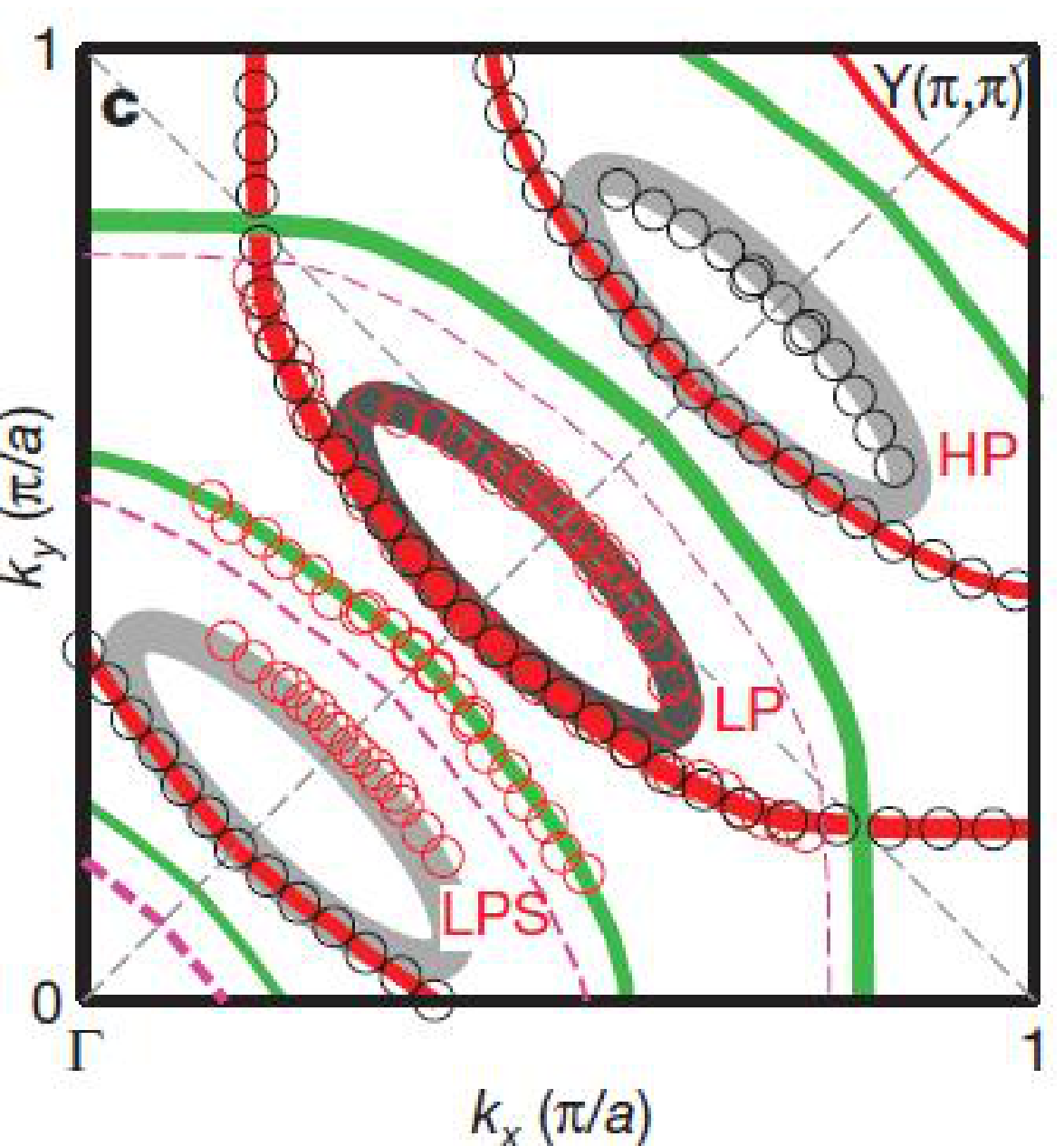}}
\caption{(a) Fermi surfaces given by the logarithmic spectral function $A(k,\omega=0)$  in the first quadrant without pairing potential. Other parameters are set to be $t =1, t' = -0.4, \mu = -1.12, kT = 0.002,  M_0 =0.7$  for a hole concentration of 0.116 and a stripe wave vector $(7\pi/8, 7\pi/8)$. The main band (MB) is given by $\epsilon_k = \mu$ and the shadow band (SB) is obtained from a reflection of MB about the fBZ boundary. (b) The Fermi surface of an under-doped ${\rm Bi_2Sr_{2-x}La_x CuO_{6+\delta}}$ sample measured by Meng {\it et al}\cite{Meng}, where the red and green curves are attributed to the umklapp bands and shadow bands.}
\end{figure}

\begin{figure}
\includegraphics[width=6cm,height=5cm]{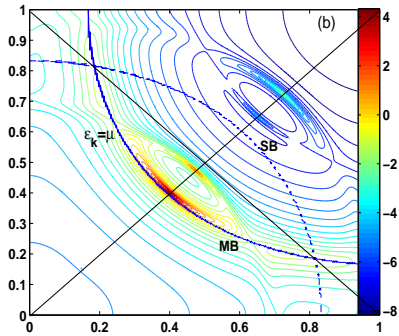}
\caption{ Fermi surfaces given by the logarithmic spectral function $A(k,\omega=0)$  in the first quadrant  of the BZ for $\Delta_d=0.2$. Other parameters see the caption in Fig.3(a).}
\end{figure}

After diagonalizing the Hamiltonian numerically the logarithmic spectral function $A(k, \omega=0)$ of the lattice is shown in Fig.2(a). It exhibits three fermi pockets on the nodal line. The central one is strong while the other two are a-few-order weaker. Surprisingly they have almost the same shape, size and position as those observed by Meng {\it et al.} in underdoped La-Bi2201(Fig.2(b))\cite{Meng}. They are apart by a wavevector of about $(0.24\pi, 0.24\pi)$ which also agrees well with the observed superstructure. The red and green curves in Fig.2(b) were attributed to the umklapp bands and shadow bands, and the curves labeled as LPS, LP, HP were thought intrinsic. It is seen that  the present computation reproduces the intrinsic excitations exactly.

In addition, strong spectrum weights still exist around the antinodal region, not coincident with the measurements. As pointed by Granath and Andersen a d-wave BCS-type coupling may remove these excitation\cite{Granath}. A d-wave pairing Hamiltonian is given by
\begin{align}
H_d &=- \sum_k ( \Delta_k C^\dagger_{k\uparrow} C^\dagger_{\bar k \downarrow}+ h.c.),\\
\Delta_k&=\Delta_d(\cos k_x - \cos k_y)/2.
\end{align}
After such a pairing potential is incorporated into the stripe model the spectrum function is slightly changed as seen in Fig.3. The central pocket and the outer one still appear  but the one close to the $\Gamma$ point does not. The most important is that the spectral weights around the antinodal region disappears, in accordance with the computation of  Granath and Andersen\cite{Granath} and the observation of Meng et al\cite{Meng}. This is a two-gap picture for the pseudogap phenomenon which goes into the temperature region above Tc.

\begin{figure}
\includegraphics[width=6cm,height=5cm]{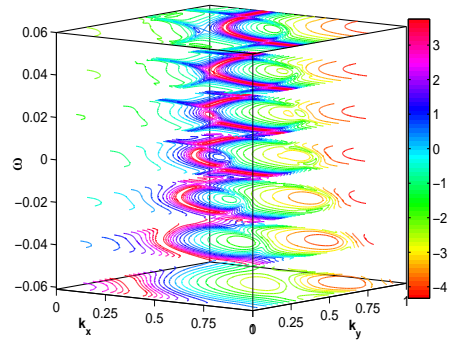}
\caption{Spectral functions $A(k,\omega)$ at low energies without pairing in the first quadrant of the BZ. Parameters see Fig.3(a). }
\end{figure}
The spectral function $A(k,\omega)$ at low energies are shown in Fig.4. It is clearly seen that the intrinsic excitations LP and HP appear below the Fermi energy and gradually move towards the $\Gamma$ point but the MB moves outwards. At the Fermi energy ($\omega=0.0$) the LP closes the Fermi surface to form the central Fermi pocket. This shows that the LP excitation is similar to the SB in the incommensurate case of the stripe phase. According to this understanding the other two pockets seen in the Fig.2(a)  are closed by the umklapp curves of the MB and the corresponding SBs. Therefore, fermi pockets in cuprates can be fully attributed to the short-range diagonal stripe phase structure.

\begin{figure}
\subfigure{\includegraphics[width=4.2cm,height=3.8cm]{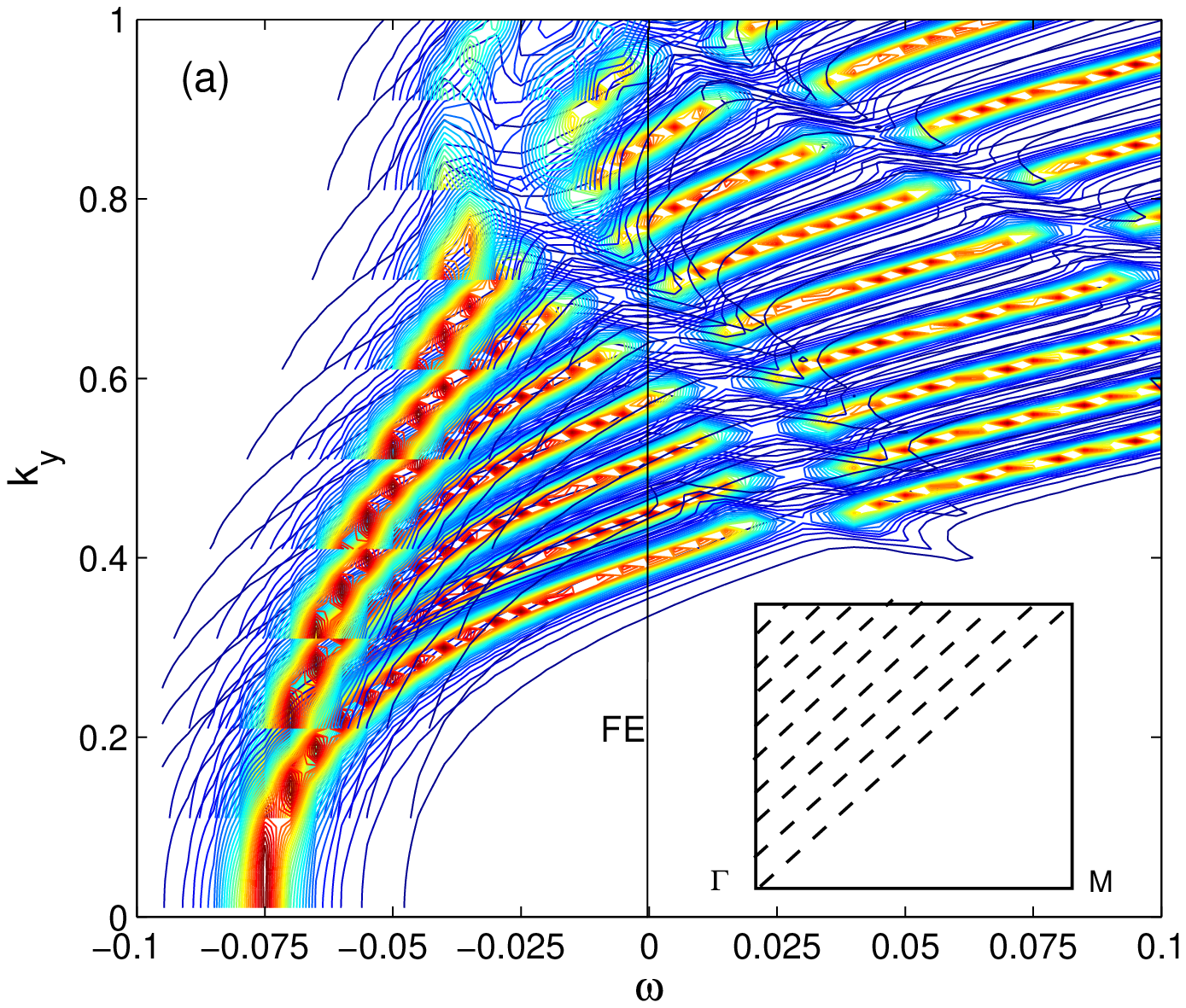}}
\subfigure{\includegraphics[width=4.2cm,height=3.8cm]{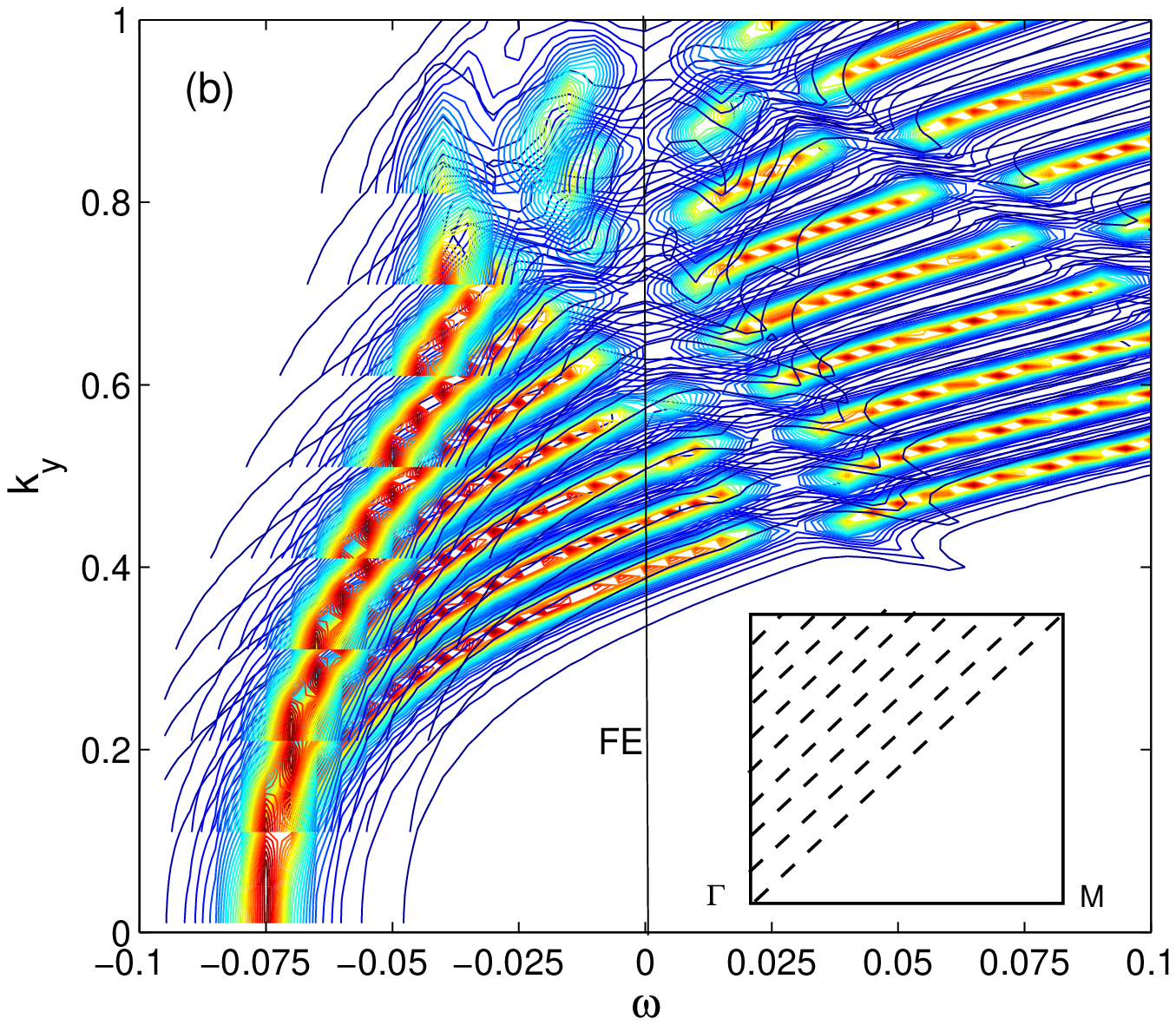}}
\caption{Spectral functions $A(k,\omega)$ at low energies along the wavevectors as shown by the dashed lines in the inset with up-to-down correspondence. (a) for $V_d$=0.0 and (b) for $V_d$=0.2.  FE denotes the Fermi energy. }\end{figure}

Fig.5 shows the comparison between the spectral functions  without and with a pairing potential. As seen in (a) the spectral function peak is cut by the diagonal stripe phase at two energies. Towards the M point $(0,\pi)$ of the BZ the two cuts gradually meet each other and spectral weight lowers significantly at M. The stripe phase opens a gap above the fermi energy ($\omega=0$) in the nodal region and below in the antinodal region.
After a d-wave pairing potential is added a new cut at the Fermi energy appears from $(0.6\pi, 0.6\pi)$ towards the antinodal region but the remaining parts of the spectral function do not change significantly. Further computation shows that the pairing potential almost keeps spectral function unchanged on the nodal line. Therefore, it is believed that the Fermi pockets in cuprates originate from the stripe phase structure but nothing to do with the d-wave pairing of electrons. This further supports the two-gap picture of the pseudogap in cuprates.

\begin{figure}
\subfigure{\includegraphics[width=4.2cm,height=3.8cm]{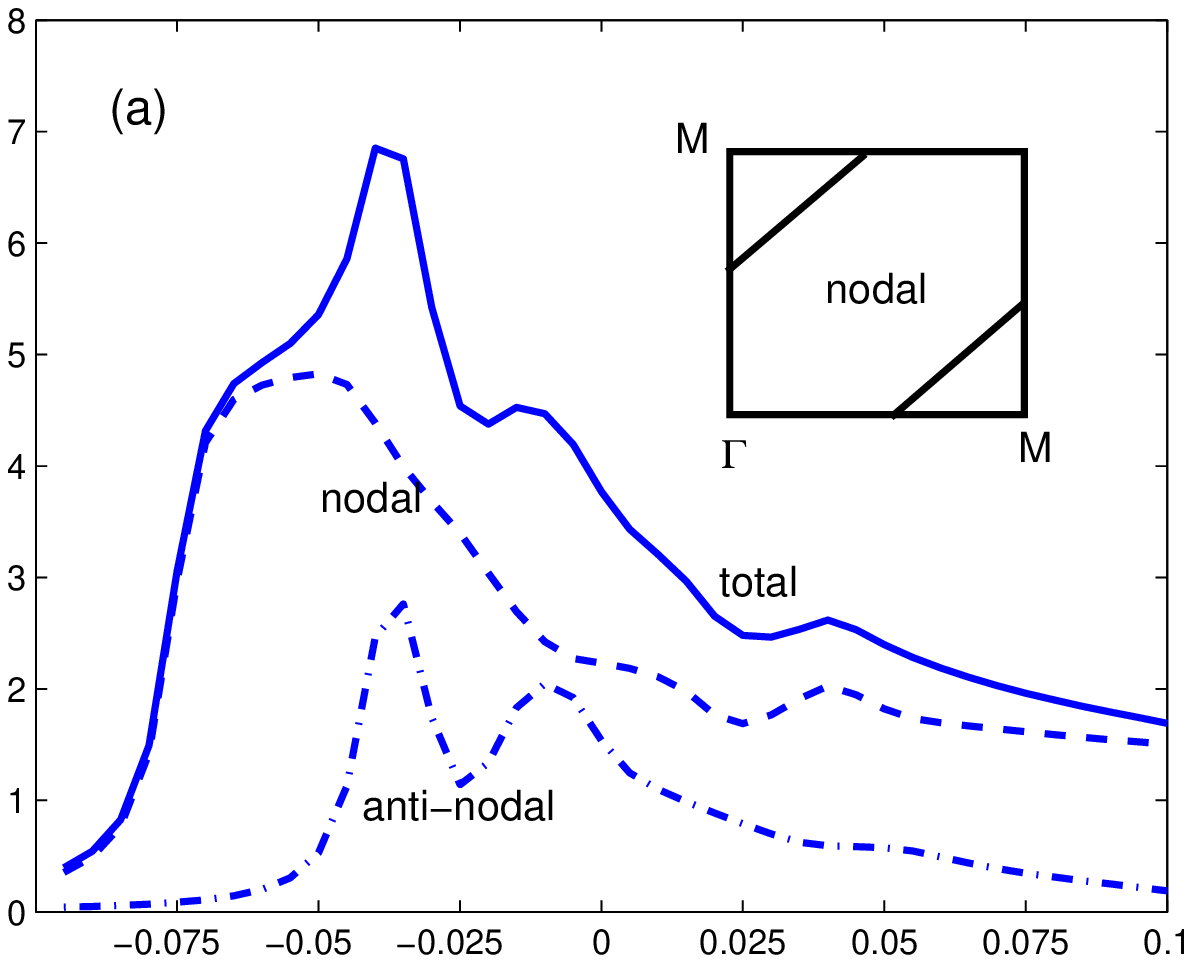}}
\subfigure{\includegraphics[width=4.2cm,height=3.8cm]{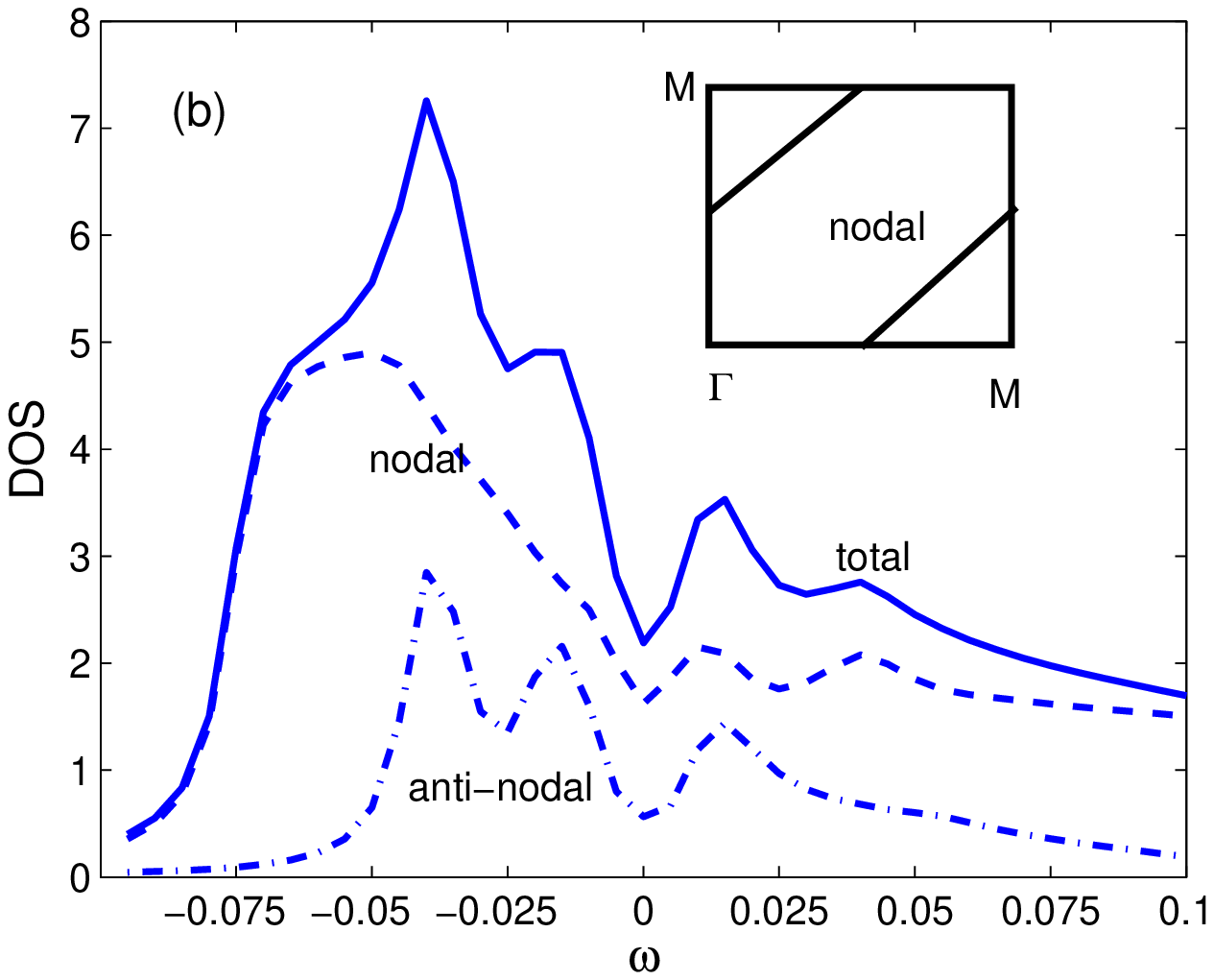}}
\caption{The density of states in the antinodal region and in the nodal region without a pairing potential (a) and with a d-wave pairing potential $V_d = 0.2$ (b).}
\end{figure}

Finally we plot the partial densities of states (DOS) in Fig.6 in the nodal region and in the antinodal region. The DOS is calculated by partially integrating the spectral function over the corresponding region, i.e. $\rho(\omega) =\sum'_k {A(k,\omega)}$. As seen in (a) without a pairing potential the partial DOS has a dip under the fermi energy in the antinodal region and a small valley above the fermi energy in the nodal region, corresponding to the cuts around the fermi energy in Fig.5(a).  With a d-wave pairing potential a clear dip appears at the fermi energy mainly in the antinodal region  as expected.

In this paper we studied the pseudogap state of cuprates in a short-range diagonal stripe phase with wave vectors $(7\pi/8, 7\pi/8)$, which reproduces accurately the positions and sizes of the fermi arc and fermi pocket and the superstructure observed by Meng et al\cite{Meng}.  The low-energy spectral function indicates that the fermi pocket results from the main band and the shadow band at the fermi energy. Above the fermi energy the shadow band gradually departs away from the main band, leaving a fermi arc. Thus we conclude that the fermi arc and fermi pocket can be fully attributed to the stripe phase but has nothing to do with pairing.  Incorporating a d-wave pairing potential into the diagonal stripe phase the spectral weight in the antinodal region is removed, leaving a clean fermi pocket in the nodal region.

This work was supported by the National Natural Science Foundation of China (Grant No.11104086) and the Natural Science Foundation of Guangdong Province of China (Grant No.S2011040001908).


\references
\bibitem{Doiron} Doiron-Leyraud, N. et al. , Nature 447, 565-568 (2007).
\bibitem{Banura} Bangura, A. et al. , Phys. Rev. Lett. 100, 047004 (2008).
\bibitem{Meng} J. Meng, Guodong Liu, Wentao Zhang {\it et al.}, Nature 462, 335(2009).
\bibitem{Tanaka} Tanaka K, Lee W S, Lu D H, et al. , Scienc,2006, 314(5807): 1910
\bibitem{Lee} Lee W S, Vishik I M, Tanaka K, et al.,  Nature, 2007, 450(7166): 81
\bibitem{Zheng} Zheng Yong, SU Gang,SCIENCE IN CHINA PRESS G, 2009: 1553 ~ 1570
\bibitem{Wen} Wen H H, Lei Shan, Xiao-Gang Wen,  et al. ,  Phys. Rev B, 2005, 72:134507.
\bibitem{Takeshi} Takeshi Kondo1, Yoichiro Hamaya, Ari D. Palczewski, et al.,  NATURE PHYSICS,  7,  2011:21
\bibitem{King}P.D. King, J.A.Rosen, W. Meevasana, et al., Phys. Rev. Lett. 106, 127005(2011).
\bibitem{Zhou} X. J. Zhou, Jiaoqiao Meng, Y. Peng, et al., arXiv: 1012.3602v1.
\bibitem{Oganesyan}Oganesyan V, ussishkin I. , Phys Rev B, 70(5):54503 (2004)
\bibitem{Agterberg} Agterberg D F, Tsunetsuga H,  Nature Phys., 2008, 4(8):639
\bibitem{Hoffman} Hoffman J E, Hudson E W, Lang K M, et al. , Science, 2002, 295(5554):466
\bibitem{Han}Han-Yong Choi and Seung Hwan Hong, Phys.Rev.B 82, 094509(2010).
\bibitem{Qiang}Qiang Han, Tao Li, and Z. D. Wang£¬Phys.Rev.B82, 052503(2010).
\bibitem{Granath} Mats Granath and Brian M. Andersen, Phys. Rev. B{\bf 81}, 024501(2010).
\bibitem{Abbamonte} P. Abbamonte et al., Nat. Phys. 1, 155 (2005).
\bibitem{Mook} H. A. Mook et al., Nature (London) 404, 729 (2000).
\bibitem{Tranquada} J. M. Tranquada, H. Woo, T. G. Perring, Nature 429, 3 (2004).
\bibitem{Hinkov} V. Hinkov, D. Haug, B. Fauqu¨¦, et al, Science 319, 597 (2008).
\bibitem{Haug} D. Haug, V. Hinkov, A. Suchaneck, et al, Phys. Rev. Lett. 103, 017001 (2009).
\bibitem{Chang} J. Chang, R. Daou, Cyril Proust, Phys. Rev. Lett. 104, 057005 (2010).
\bibitem{Seibold} G. Seibold, J. Lorenzana  and M. Grilli, PHYS. REV. B 75, 100505(R) (2007).
\bibitem{Matthias} Matthias Vojta, Thomas Vojta  and Ribhu K. Kaul, Phys. Rev. Lett. 97, 097001 (2006)
\bibitem{Yang} Yang H B, Rameau J D, Johnson P D, et al., Nature 2008, 456(7218) 77.
\bibitem{Marshall} Marshall D S, Dessau D S , Loeser AG, et al, Phys. Rev. Lett. 76, 4841(1996).
\end{document}